\documentclass[twoside]{article}
\usepackage{fleqn,espcrc2}
\usepackage{graphicx}
\usepackage{here}

\newcommand{\AmS}{{\protect\the\textfont2
    A\kern-.1667em\lower.5ex\hbox{M}\kern-.125emS}}										
\def\beq{\begin{equation}}
\def\eeq{\end{equation}}
\def\bea{\begin{eqnarray}}
\def\eea{\end{eqnarray}}
\def\bq{\begin{quote}}
\def\eq{\end{quote}}

\def\nnb{\nonumber}
\def\ga{\left(}
\def\dr{\right)}

\def\rar{\rightarrow}
\def\lrar{\Longrightarrow}
																
\def\nnb{\nonumber}
\def\la{\langle}
\def\ra{\rangle}
\def\nin{\noindent}
\def\ba{\vspace*{-0.2cm}\begin{array}}
\def\ea{\end{array}\vspace*{-0.2cm}}

\def\als{\alpha_s}

\def\gg2{ \la\alpha_s G^2 \ra}
\def\gg3{g^3f_{abc}\la G^aG^bG^c \ra}
\def\ggg4{\la\als^2G^4\ra}

\title
{\bf{\boldmath
{\Large $1^{-+}$ light exotic mesons in QCD 
} }}

\author
{ 
Stephan Narison\thanks{Email: snarison@yahoo.fr}  \address {\footnotesize Laboratoire de Physique Th\'eorique et Astroparticules, CNRS - IN2P3 \& Universit\'e
de Montpellier II, Case 070, Place Eug\`ene Bataillon, 34095 - Montpellier Cedex 05, France},
}

\begin{document}

\pagestyle{myheadings}
\markright{ }
\begin{abstract}
\noindent
We systematically re-examine the extraction of the masses and couplings of the $1^{-+}$ 
hybrid, four-quark and molecule mesons from QCD spectral sum rules (QSSR). To NLO for the perturbative and power corrections, the hybrid mass is $M_H=1.81(6)$ GeV and $M_H\leq 2.2(2)$ GeV from the positivity of the spectral function. In the same way , but  to LO, the four-quark state mass is $M_{4q}= 1.70(4)$ GeV and $M_{4q}\leq 2.4(1)$ GeV, while the molecule mass is about $1.3(1)$ GeV. The observed $\pi_1(1400)$ and $\pi_1(1600)$ might be explained by a two-component mixing with the set of input masses $(1.2\sim 1.3~;~1.70\sim 1.74)$ GeV and with a mixing angle $\theta\simeq -(11.7\pm 2.2)^0$, which slightly favours a molecule/four-quark mixing, and which eventually suggests that the $\pi_1(2015)$ is mostly an hybrid meson. 
Isospin and non-exotic partners of the previous states and some of their radial excitations are also expected to be found in the energy region around 2 GeV. Further  tests of this phenomenological scenario are required.
\end{abstract}
\maketitle
\section{Introduction}
\nin
Light hybrid mesons with the exotic quantum number $1^{-+}$ and with characterisitc decays into $\rho\pi$, $b_1\pi$ and $\eta'\pi$ have been expected to be easily found in hadronic experiments. There are indeed two good experimental candidates  $\pi_1(1400),~\pi_1(1600)$ \cite{PDG},  but their masses are lower than the lattice predictions in the quenched approximation  \cite{LATT1} and with two dynamical quarks (2q) \cite{LATT2}:
\beq
M_H\vert_{\rm quenched}\simeq 1.9~{\rm GeV}~, ~~~~ M_H\vert_{\rm 2q}\simeq 2.2(2)~{\rm GeV}~.
\label{eq:latt}
\eeq
 First studied in QCD in \cite{HYG,HYG1}, lowest order (LO) results from QCD spectral sum rules (QSSR)\cite{SNB} \`a la SVZ \cite{SVZ}  have been obtained in \cite{HYG2,LNP,SNH}:
\beq
M_H\vert_{\rm LO}\simeq (1.6\sim 2.1)~{\rm GeV}~,
\eeq
which, after the inclusion of radiative corrections (partly obtained in \cite{CHET} and completed in \cite{KORNER}) lead to the upper bound and estimate in units of GeV: 
\beq
M_H\vert_{\rm NLO}\leq (1.9\sim 2.0)~,~~~~
M_H\vert_{\rm NLO}= (1.6\sim 1.7)~.
\label{eq:qssr}
\eeq
Though the QSSR result tends to favour the hybrid assignement of the $\pi_1$(1600) meson, as quoted in \cite{PDG}, it is
important to understand the discrepancy of the above QSSR results with the ones from the lattice
and to a lesser extent with the one in \cite{YANG} where it is claimed that the inclusion of the operator anomalous dimension obtained in \cite{CHET} decreases substantially the previous result in Eq. (\ref{eq:qssr}) down to 1.2 GeV, though a such effect disappear to LO, and a similar one in the tensor meson has lead to negligible corrections \cite{BAGAN}.  
 One can also notice that  \cite{KORNER} use as input the value of the four-quark condensate from vacuum saturation which has been shown from different channels and different groups to be violated  by a factor 2-3 \cite{SNB,LNT,fesr,SNTAU,SNI,JAMI2}.  In the following, we revisit the existing NLO results by paying attention on the influence of different QCD input parameters used in the analysis. We shall also check the effect of the operator anomalous dimension and study the new effect of the tachyonic gluon mass introduced in \cite{CHET,CNZ,ZAK} but overlooked in \cite{KORNER}. For a more robust estimate, we shall consider results both from finite energy (FESR) and Laplace (LSR) sum rules. 
We complement our studies by reexamining the predictions obtained  for the $1^{-+}$ light
four-quark \cite{ZHU} and molecule \cite{ZHANG} states and conclude the paper by a phenomenological discussion on the possible nature of the observed $\pi_1(1400)$, $\pi_1(1600)$ and $\pi_1(2015)$ mesons.
 \section{QCD spectral sum rules (QSSR)}
\subsection*{Description of the method}
\nin
Since its discovery in 1979 \cite{SVZ}, QSSR has proved to be a
powerful method for understanding the hadronic properties in terms of the
fundamental QCD parameters such as the QCD coupling $\alpha_s$, the (running)
quark masses and the quark and/or gluon QCD vacuum condensates \cite{SNB}.
In practice (like also the lattice), one starts the analysis {}from the
two-point correlator
(standard notations):
\bea
\label{corr}
\Pi^{\mu\nu}_{V/A}(q^2)& \equiv& i \int d^4x ~e^{iqx} 
\la 0\vert {\cal T}
{\cal O}_{V/A}^\mu (x)
\ga {\cal O}_{V/A}^\nu (0)\dr ^\dagger \vert 0 \ra \nnb\\
&=&-\ga
g^{\mu\nu}q^2-q^\mu q^\nu\dr\Pi^{(1)}_{V/A}(q^2)\nnb\\
&&+q^\mu
q^\nu\Pi^{(0)}_{V/A}(q^2),
\eea
built {}from the hadronic local currents ${\cal O}^{V/A}_\mu (x)$:
\bea
\label{oper}
{\cal O}_{V}^\mu (x)&\equiv& :g\bar \psi_i\lambda_a \gamma_\nu \psi_j
G^{\mu\nu}_a: \ ,\nnb\\
\quad {\cal O}_A^\mu (x)&\equiv & :g\bar \psi_i\lambda_a
\gamma_\nu \gamma_5\psi_j G^{\mu\nu}_a:
\eea
which select the specific quantum numbers
of the hybrid mesons; A and V refer respectively to the vector and
axial-vector currents.
The invariant $\Pi^{(1)}$ and $\Pi^{(0)}$ refer to the spin one and zero
mesons.
One exploits, in the sum rule approaches, the analyticity property of the
correlator which obeys the well-known K\"allen--Lehmann dispersion relation:
\beq
\Pi_{V/A}^{(1,0)} (q^2) =
\int_{0}^{\infty} \frac{dt}{t-q^2-i\epsilon}
~\frac{1}{\pi}~\mbox{Im} ~ \Pi_{V/A}^{(1,0)}  + ...
\eeq
where ... represent subtraction terms which are
polynomials in the $q^2$-variable. In this way, the $sum~rule$
expresses in a clear way the {\it duality} between the integral involving the
spectral function Im$ \Pi_{V/A}^{(1,0)}(t)$ (which can be measured
experimentally),
and the full correlator $\Pi_{V/A}^{(1,0)}(q^2)$. The latter
can be calculated directly in 
QCD in the Euclidean space-time using perturbation theory (provided  that
$-q^2+m^2$ ($m$ being the running quark mass) is much greater than $\Lambda^2$),
and the Wilson
expansion in terms of the increasing dimensions of the quark and/or gluon
condensates which
 simulate the non-perturbative effects of QCD.
\subsection*{The SVZ expansion and beyond}\label{sec:qcdparam}
\nin
Using the Operator Product Expansion (OPE) \cite{SVZ}, the two-point
correlator reads for $m=0$:
$$
\Pi^{(1,0)}_{V/A}(q^2)
\simeq \sum_{D=0,2,...}\frac{1}{\ga q^2 \dr^{D/2}}
\sum_{dim O=D} C(q^2,\nu)\la {\cal O}(\nu)\ra~,
$$
where $\nu$ is an arbitrary scale that separates the long- and
short-distance dynamics; $C$ are the Wilson coefficients calculable
in perturbative QCD by means of Feynman diagrams techniques; $\la {\cal
O}(\nu)\ra$
are the quark and/or gluon condensates of dimension $D$.
In the massless quark limit, one may expect
the absence of terms of dimension 2 due to gauge invariance. However,
it has been
emphasized recently \cite{ZAK} that  the resummation of the large order
terms of the
perturbative series can
be mimiced by the effect of a tachyonic gluon mass $\lambda$ which
generates an extra $1/q^2$ term not present in
the original OPE. This short distance mass has been estimated from the
$e^+e^-$ data \cite{SNI,CNZ} and pion sum rule \cite{CNZ} to be:
\beq
\label{lamb}
\frac{\alpha_s}{\pi}\lambda^2\simeq -(0.07\pm 0.03) ~\rm{ GeV}^2.
\eeq
In addition to Eq.~(\ref{lamb}), the strengths of the vacuum condensates having
dimensions $D\leq 6$ are also under
good control, namely:
\begin{itemize}
\item $\la\alpha_s G^2\ra \simeq (6.8\pm1.3)10^{-2}~\rm{GeV}^4$ {}from
sum rules of $ e^+e^-\rar I=1~\rm{hadrons}$ \cite{SNI} 
and {heavy quarkonia} \cite{BB,SNHeavy,YNDU};
\item $g\la\bar{\psi}G\psi\ra\equiv g\la\bar{\psi}\lambda_a\sigma^{\mu\nu}G^a_{\mu\nu}\psi\ra\simeq
2\times (0.8\pm
0.1)~{\rm GeV}^2\la\bar \psi\psi\ra,$ {}from
the baryons \cite{HEID,JAMI2} and
the heavy-light mesons \cite{SNhl} systems;
\item $\alpha_s  \la\bar \psi\psi\ra^2\simeq
(4.5\pm 0.3) \times 10^{-4}~\rm{ GeV}^6$ {}from
$~e^+e^-\rar I=1~ \rm{hadrons}$ \cite{SNI} and $\tau$-decay data \cite{SNTAU}, where a deviation from the vacuum saturation estimate has been noticed from different studies \cite{SNB} and \cite{LNT,fesr,SNTAU,SNI};
\item $g^3\la G^3\ra\simeq (1.2\pm 0.2)~\rm{GeV}^2\la\alpha_s G^2\ra $ 
{}from { dilute gaz instantons}~\cite{NSVZ} and lattice calculations \cite{DIGI2}.
\end{itemize}
In the numerics, we shall use the value of the QCD scale:
\beq
\Lambda_3= (353\pm 15)~{\rm MeV}~,
\eeq
deduced recently to 4-loops from the value of $\alpha_s(M_\tau)=0.3249(80)$ from $\tau$-decay \cite{SNTAU}. 
\subsection*{Spectral function}
\nin
In the absence of the complete data, the spectral function is often parametrized using the ``na\"{\i}ve" duality ansatz:
\bea
\frac{1}{\pi}~\mbox{Im}  \Pi^{(1,0)}_{V/A}(t)&\simeq& 2M_H^{4}f_H^2 \delta
(t-M_H^2)\nnb\\
&+&
 \rm{``QCD
~continuum"}
\times \theta(t-t_c)~,
\eea
which has been tested \cite{SNB} using $e^+e^-$ and $\tau$-decay data, to
give a good description of the
spectral integral in the sum rule analysis even in the case of the broad $\sigma$ and $K^*_0$ states \cite{VENEZIA}; $f_H$ 
(an analogue to $f_\pi$) is
the hadron's coupling to the current;
while $t_c$ is the QCD continuum's threshold.
\subsection*{Sum rules and optimization procedure}
\nin
Among the different sum rules discussed in the literature \cite{SNB}, we
shall be
concerned with the following {Laplace sum rule (LSR)} and its ratio
\cite{SVZ,NR,BB} :
\bea\label{usr}
{\cal L}^{(1,0)}_n(\tau)
&=& \int_{0}^{\infty} {dt}~t^n~\mbox{exp}(-t\tau)
~\frac{1}{\pi}~\mbox{Im} \Pi_{V/A}^{(1,0)}(t)~,\nnb\\{\cal R}_{n}(\tau)
&\equiv& -\frac{d}{d\tau} \log {{\cal
L}_n}~,~~~~~~~(n\geq 0) \ .
\eea
The advantage of the Laplace sum
rules with respect to the previous dispersion relation is the
presence of the exponential weight factor which enhances the
contribution of the lowest resonance and low-energy region
accessible experimentally. For the QCD side, this procedure has
eliminated the ambiguity carried by subtraction constants,
arbitrary polynomial
in $q^2$, and has improved the convergence of
the OPE by the presence of the factorial dumping factor for each
condensates of given dimensions.
The ratio of the sum rules is a useful quantity to work with,
in the determination of the resonance mass, as it is equal to the
meson mass squared, in the usual duality ansatz parametrization.
As one can notice, there are ``a priori" two free external parameters $(\tau,
t_c)$ in the analysis. The optimized result will be (in principle) insensitive
to their variations. In some cases, the $t_c$-stability is not reached due
to the
too na\"{\i}ve parametrization of the spectral function.
In order to restore 
the $t_c$-stability of the
results 
one can 
fix  the
$t_c$-values by the help of FESR (local duality) \cite{Chet,fesr}:
\bea\label{fesr}
{\cal M}^{(1,0)}_n
&=& \int_{0}^{t_c} {dt}~t^n
~\frac{1}{\pi}~\mbox{Im} \Pi_{V/A}^{(1,0)}(t)~,\nnb\\
R_n(t_c)
&\equiv&{ \int_{0}^{t_c} {dt}~t^{n+1}
~\frac{1}{\pi}~\mbox{Im} \Pi_{V/A}^{(1,0)}(t)\over
\int_{0}^{t_c} {dt}~t^n
~\frac{1}{\pi}~\mbox{Im} \Pi_{V/A}^{(1,0)}(t)} \simeq M_H^2~.
\eea
The results discussed in the next section will satisfy these stability
criteria.
\section{The hybrid two-point function in QCD}
\nin
A QCD analysis of the two-point function have been
done in the past by different groups \cite{HYG,HYG1}, where (unfortunately)
the non-trivial QCD expressions were wrong
leading to some controversial predictions \cite{SNB}. 
In this paper, we extend the analysis by taking into account
the non-trivial $\alpha_s$ correction and the
effect of the new $1/q^2$ term not taken into account into the SVZ expansion.
The corrected QCD expressions of the correlator are given in
\cite{HYG2,LNP,SNB} to lowest order of perturbative QCD
but including the contributions of the condensates of dimensions lower than or
equal to six.
The new terms appearing
in the OPE are presented in the following \footnote{
We neglect some possible 
mixings  with operators containing more $\gamma$-matrices
like 
$ 
g\bar \psi_i\lambda_a \gamma_\mu \sigma_{\nu\lambda} \psi_j
G^{\nu\lambda}_a$, which is expected to be small.}:
\begin{itemize}
\item The perturbative QCD expression including the NLO radiative
corrections reads \cite{CHET}:
\bea
\mbox{Im}\Pi_{V/A}^{(1)}\vert_{pert}&=&\frac{\alpha_s}{60\pi^2}t^2
\Bigg{[} 1+\frac{\alpha_s}{\pi}\Big{[}\frac{121}{16}\nnb\\&-&\frac{257}{360}n_f
+\ga
\frac{35}{36}-\frac{n_f}{6}\dr
\log{\frac{\nu^2}{t}}\Big{]}\Bigg{]}\nnb\\
\mbox{Im}\Pi_{V/A}^{(0)}\vert_{pert}&=&\frac{\alpha_s}{120\pi^2}t^2
\Bigg{[} 1+\frac{\alpha_s}{\pi}\Big{[}\frac{1997}{432}-\frac{167}{360}n_f
\nnb\\&+&\ga
\frac{35}{36}-\frac{n_f}{6}\dr
\log{\frac{\nu^2}{t}}\Big{]}\Bigg{]}
\eea
\item The anomalous dimension of the current can be easily deduced to 
be \cite{CHET}:
\beq
\gamma\equiv \nu \frac{\mathrm{d}}{\mathrm{d} \nu} {\cal O}_{V}^\mu = 
\Big{[} \gamma_1\equiv -\frac{16}{9}\Big{]} \frac{\alpha_s}{\pi} {\cal O}_{V}^\mu
{}.
\label{anom.dim}
\eeq
\item The lowest order correction due to the (short distance)
tachyonic gluon mass reads:
\bea
\mbox{Im}\Pi_{V/A}^{(1)}(t)_{\lambda}&=&-\frac{\alpha_s}{60\pi^2}
\frac{35}{4}\lambda^2t\nnb\\
\mbox{Im}\Pi_{V/A}^{(0)}(t)_{\lambda}&=&\frac{\alpha_s}{120\pi^2}
\frac{15}{2}\lambda^2t
\eea
\item The  contributions of the dimension-four condensates
reads in the limit $m^2=0$ to LO \cite{HYG2,LNP,SNB} and to NLO \cite{KORNER}:
\bea
\mbox{Im}\Pi_{V}^{(1)}\vert_{4}&=&\frac{1}{9\pi}
\Bigg{[} \alpha_s\la G^2\ra 
\Big{[}1-{145\over 72}{ \alpha_s\over\pi}\nnb\\
&+&{8\over 9}{ \alpha_s\over\pi}\log{\frac{t}{\nu^2}}\Big{]}
+8\alpha_s m\la\bar\psi\psi\ra\Bigg{]}\nnb\\
\mbox{Im}\Pi_{A}^{(0)}\vert_{4}&=&-\frac{1}{6\pi}\Bigg{[} 
\alpha_s\la G^2\ra
\Big{[}1-{209\over 72}{ \alpha_s\over\pi}\nnb\\
&+&{8\over 9}{ \alpha_s\over\pi}\log{\frac{t}{\nu^2}}\Big{]}
-8\alpha_s
m\la\bar\psi\psi\ra\Bigg{]}~,
\eea
where $a_s\equiv \alpha_s/\pi$ and $\la\bar\psi\psi\ra\equiv  \la\bar\psi_u\psi_u\ra\simeq \la\bar\psi_d\psi_d\ra$.
\item To leading order in $\alpha_s$, the  contributions of the dimension-six gluon and mixed condensates
read \cite{LNP,SNB}:
\bea
\Pi_{V}^{(1)}\vert_{6}&=&\frac{1}{48\pi^2q^2}\Big{[}g^3
\la G^3\ra
-\frac{83}{9}{\alpha_s}
m g\la\bar\psi G\psi\ra\Big{]}\nnb\\
\Pi_{A}^{(0)}\vert_{6}&=&\frac{11}{18}
\frac{\alpha_s}{\pi}\frac{1}{q^2}m 
g\la\bar\psi G \psi \ra \log{-\frac{q^2}{\nu^2}}~,
\eea
where one can notice the miraculous cancellation of the $\log$-coefficient
of the dimension-six condensates for $\Pi_{V}^{(1)}$.
\item The four-quark condensate contributions including radiative corrections
read for $n$ flavours \cite{KORNER}:
\bea
\Pi_{V}^{(1)}\vert_{6}&=&\frac{1}{q^2}\frac{16}{9}{\alpha_s}
\la\bar \psi\psi\ra^2\Bigg{[}1+{1\over 18}\ga {91\over 6}-{5}n_f\dr{\alpha_s\over\pi}\nnb\\
&+&{1\over 6} \ga {11\over 12}+{n}\dr {\alpha_s\over\pi}  \log{-\frac{q^2}{\nu^2}}\Bigg{]}~.
\eea
\item The contribution of the four-quark condensate
in the (pseudo)scalar channels vanishes to leading order in $\alpha_s$ and starts
at order $\alpha_s^2\la\bar\psi\psi\ra^2$. In the scalar channel, it reads \cite{KORNER}:
\bea
\Pi_{V}^{(0)}\vert_{6}&=&\frac{1}{q^2}\frac{16}{3}{\alpha_s}
\la\bar \psi\psi\ra^2\Bigg{[}{1\over 3}\ga {14\over 3}+{n_f\over 2}\dr {\alpha_s\over\pi}\nnb\\
&-&{1\over 12} \ga {53\over 12}+{n_f}\dr {\alpha_s\over\pi}  \log{-\frac{q^2}{\nu^2}}\Bigg{]}~.
\eea
\end{itemize}
\section{Upper bound on  $M_H(1^{-+})$  from LSR}
\nin
Using the positivity of the spectral function, we can deduce an upper bound on $M_{H}$ from
${\cal R}_n(\tau):~n=0,1$. We show the result in Fig. \ref{fig:usr} for the $1^{-+}$ channel. The result is given in the domain limited by the two full (red) curves from ${\cal R}_0(\tau)$ and by the two green curves (dashed) from ${\cal R}_1(\tau)$ spanned by the two extremal values of the set of QCD
parameters given in Table \ref{tab:svz}. 
\begin{center}
{\begin{figure}
{\includegraphics[width=6cm]{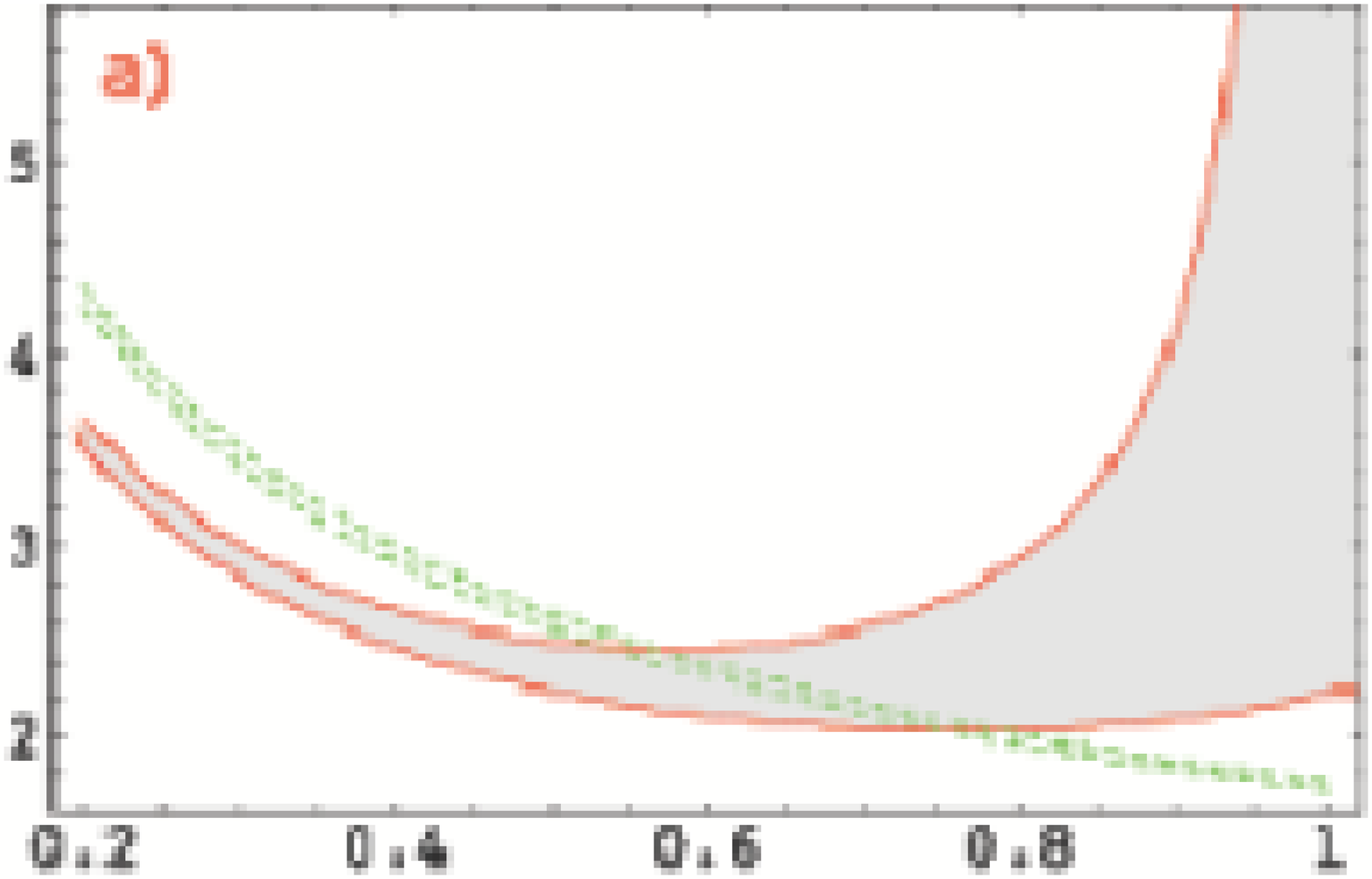}}
{\includegraphics[width=6cm]{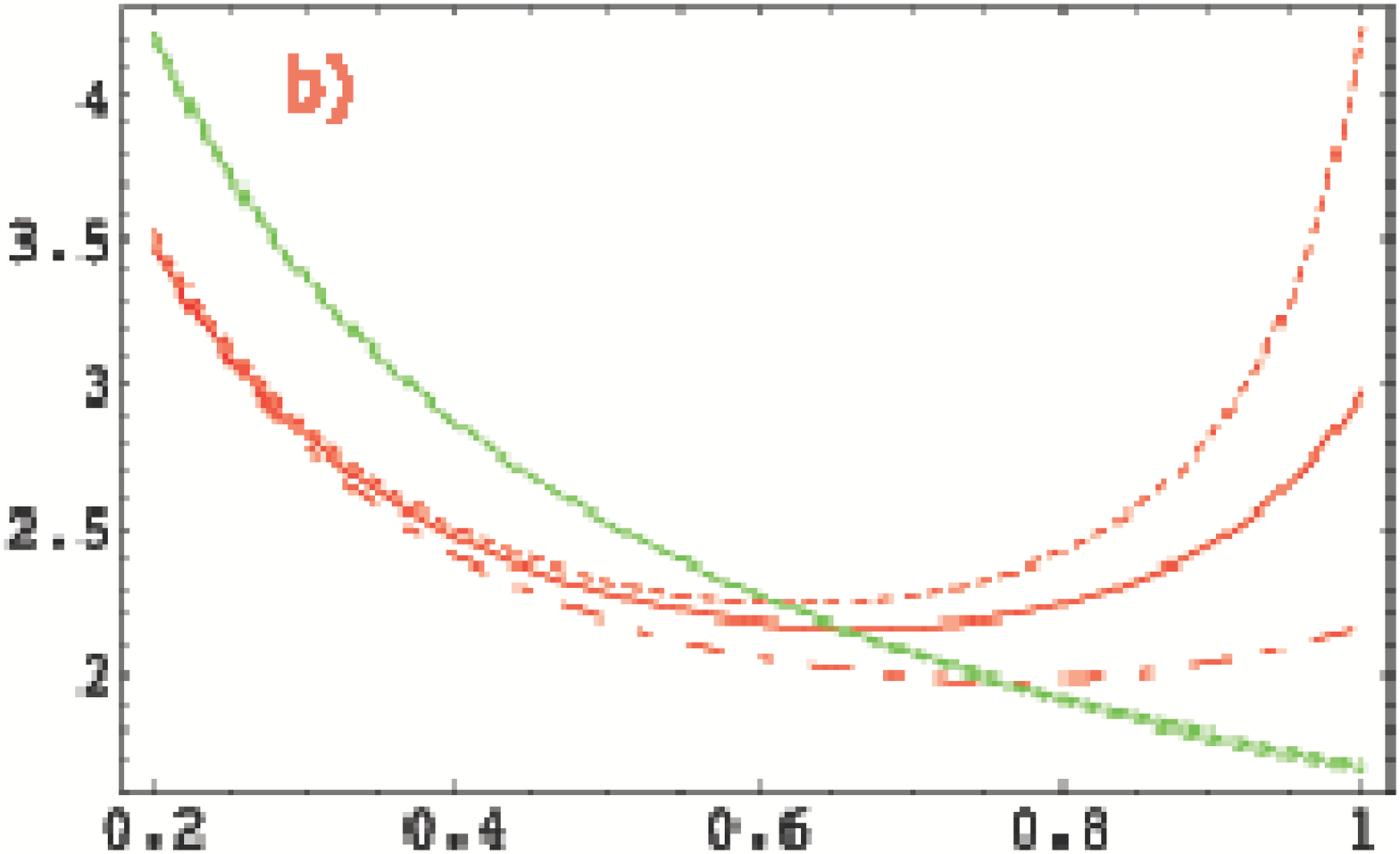}}
 \caption{\scriptsize 
 a) Domain of upper bound on $M_H$ in GeV versus the LSR variable $\tau$ in GeV$^{-2}$ for two extremal values of the Sets of Power Corrections in Table \ref{tab:svz} from ${\cal R}_0(\tau)$ (red: continuous) and ${\cal R}_1(\tau)$ (green: dashed); b) Effect of radiative corrections on the central value of the bound of $M_H$ from ${\cal R}_0(\tau)$ red: : LO (dashed-dotted); NLO condensate+LO perturbative (dashed) ; Total NLO  (continuous). Green curves are from ${\cal R}_1(\tau)$.
 }
\label{fig:usr}
\end{figure}}
\end{center}
\vspace*{-0.5cm}
{\scriptsize
\begin{table}[H]
\setlength{\tabcolsep}{0.35pc}
 \caption{\scriptsize    Value of the upper bound  on $M_H$  in GeV at the minima of ${\cal R}_0(\tau)$ and at its intersection with ${\cal R}_1(\tau)$ for two extremal values of the Sets of Power Corrections in units of GeV$^d$ ($d$ is their dimension).
    }
\begin{tabular}{lcccc}
&\\
\hline
\hline
\\
 \\ Power Corrections &${(\alpha_s/\pi)}\lambda^2$&$ \alpha_s\la G^2\ra$  &$\alpha_s\la\bar\psi\psi\ra^2\times 10^4$&$M_{\pi_1} $  \\
\\
\hline
\hline
\\
Set 1&-0.04&0.055&4.8&$\leq 2.0$ \\
&\\
Set 2&-0.10&0.081&4.2&$\leq 2.4$ \\
&\\
\hline
\hline
\end{tabular}
\label{tab:svz}
\end{table}
}
\vspace*{-0.5cm}
\nin
One can note that the prediction increases by 100 MeV each when $|\lambda^2|$ and $\alpha_s\la G^2\ra$ decrease from their central value to the one allowed in the range given in Section 2, while   it increases by 25 MeV  when $10^4\alpha_s\la\bar\psi\psi\ra^2$ increases from 4.5 to 4.8. The effects of the variation of $\Lambda$ and $M_0^2$ within the range in Section 2 are invisible. At the stability points (in the sum rule variable $\tau$) of the continuous (red) curves which are also the intersections with the dashed (green) curves, one can deduce:
\beq
M_H(1^{-+})\leq (2.2\pm 0.2 )~{\rm GeV}~.
\eeq
This value confirms and improves previous results in \cite{LNP,CHET,KORNER}. The result $(2.2\pm 0.2)$ GeV from lattice with two dynamical quarks though still consistent with this bound is on its boarder. 
We show in Fig \ref{fig:usr} the effects of radiative corrections on the results. One can see that the ones due to the condensates give larger effects and increase the bound from 2.0 (LO)  to 2.25 GeV, while the ones due to the perturbative terms decrease the mass prediction by 0.05 MeV, where the $\alpha_s$-correction and the one induced by the anomalous dimension act in the opposite directions. The ratio ${\cal R}_1$ is amost unaffected by the radiative corrections [dotted (green) curves on top of each others].
\section{QSSR predictions of $f_H$ and $M_H $ for $1^{-+}$}
\nin
Following Ref. \cite{BAGAN}, we introduce the RGI coupling $\hat f_H$ defined as:
\beq
f_H(\nu)={\hat f_H\over \ga\log{\nu/\Lambda}\dr^{\gamma_1/-\beta_1}}~,
\eeq
with $\gamma_1$ the anomalous dimension in Eq. (\ref{anom.dim}) and $-\beta_1=1/2(11-2n_f/3)$ the first coefficient of the $\beta$ function for $n_f$ flavours.
\begin{center}
{\begin{figure}[H]
{\includegraphics[width=6cm]{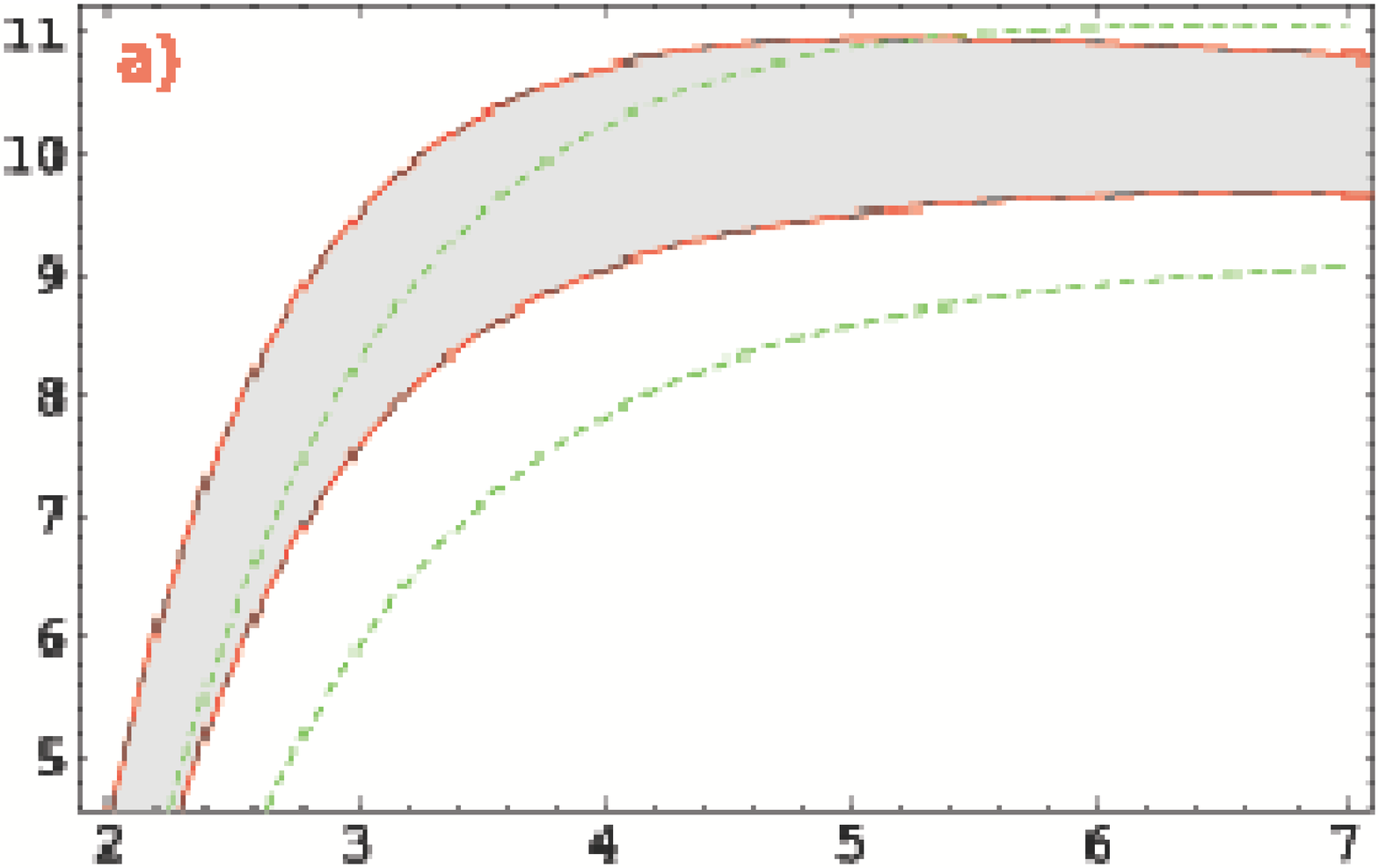}}
{\includegraphics[width=6cm]{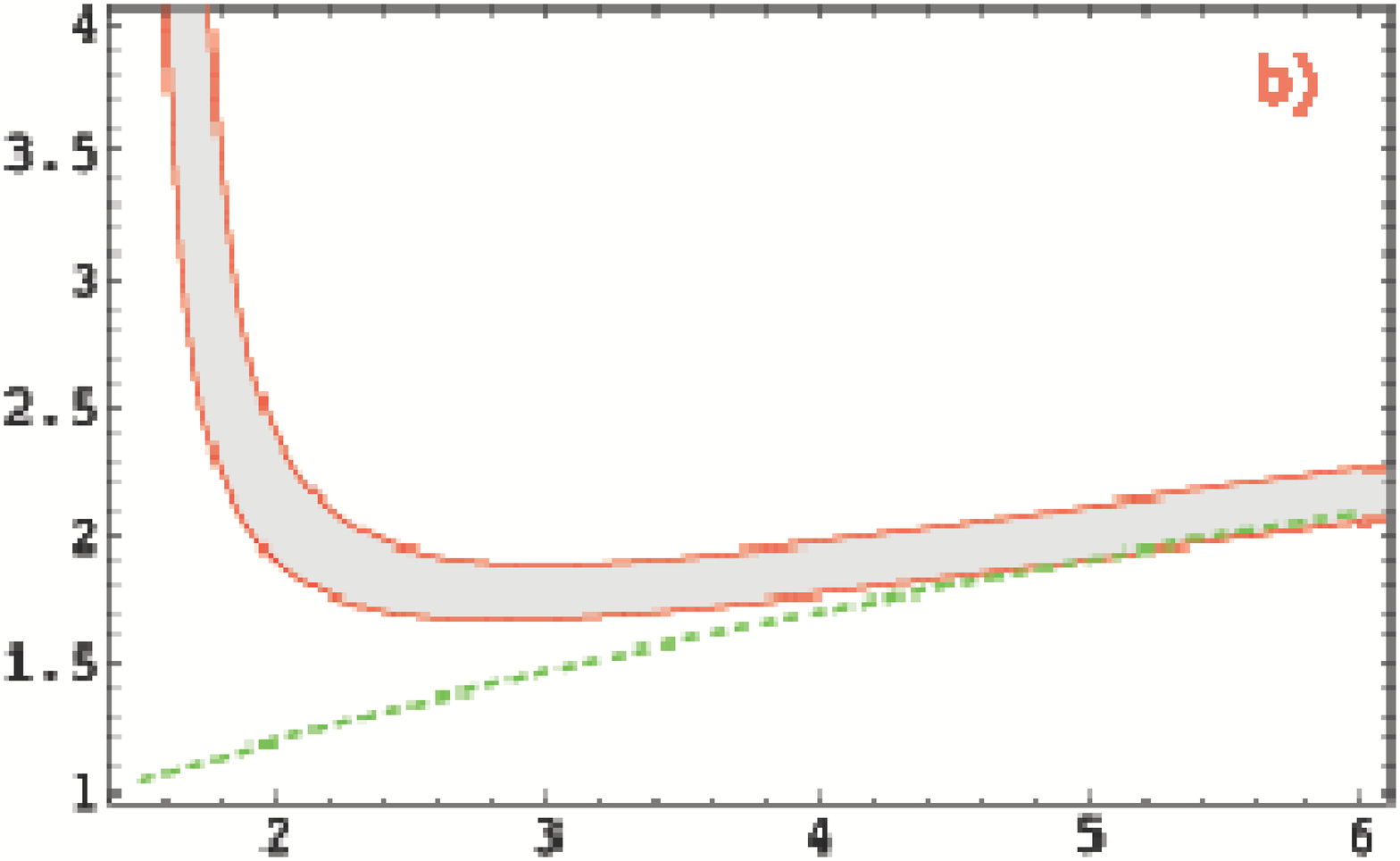}}
 \caption{\scriptsize 
a) Domain of FESR predictions versus the QCD continuum threshold $t_c$ using the set of power correction in Table \ref{tab:svz} for  $f_H$ from ${ \cal M}_0$ (red: continuous) and ${ \cal M}_1$ (black: dashed (on top of the red curve)) and ${ \cal M}_2$ (green: dotted); b) The same as for $f_H$ but for $M_H$: from $R_0(t_c)$ (red: continuous); from $R_1(t_c)$ (green: dotted).
 }
\label{fig:fesr}
\end{figure}}
\end{center}
\vspace*{-0.5cm}
Using FESR, we show  the predictions for the decay constant in Fig. \ref{fig:fesr}a) and the mass of the $1^{-+}$ in in Fig. \ref{fig:fesr}b).
Using the stability criterion on the variation  of $t_c$, and the intersection of the predictions of different moments,  and the Sets of Power Corrections in Table \ref{tab:svz}, one obtains the optimal value for the $1^{-+}$:
\bea
\hat f_H&=& 9.6(1.4)~{\rm MeV}:t_c=(4.4\sim 5.2)~{\rm GeV}^2\nnb\\
M_H&=&1.80(10)~{\rm GeV}:t_c=(2.8\sim 5.0)~{\rm GeV}^2.
\eea
\begin{center}
{\begin{figure}
{\includegraphics[width=6cm]{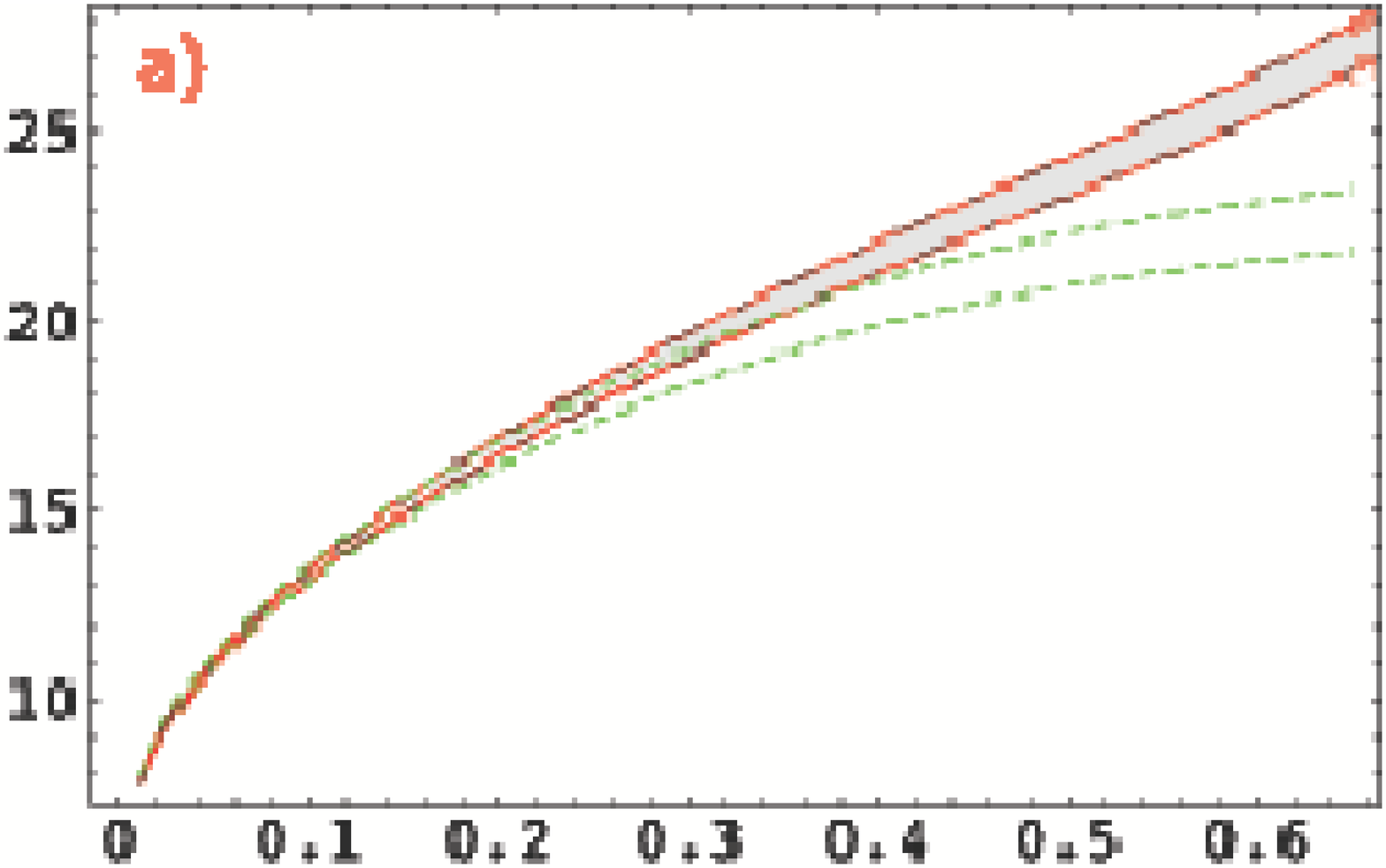}}
{\includegraphics[width=6cm]{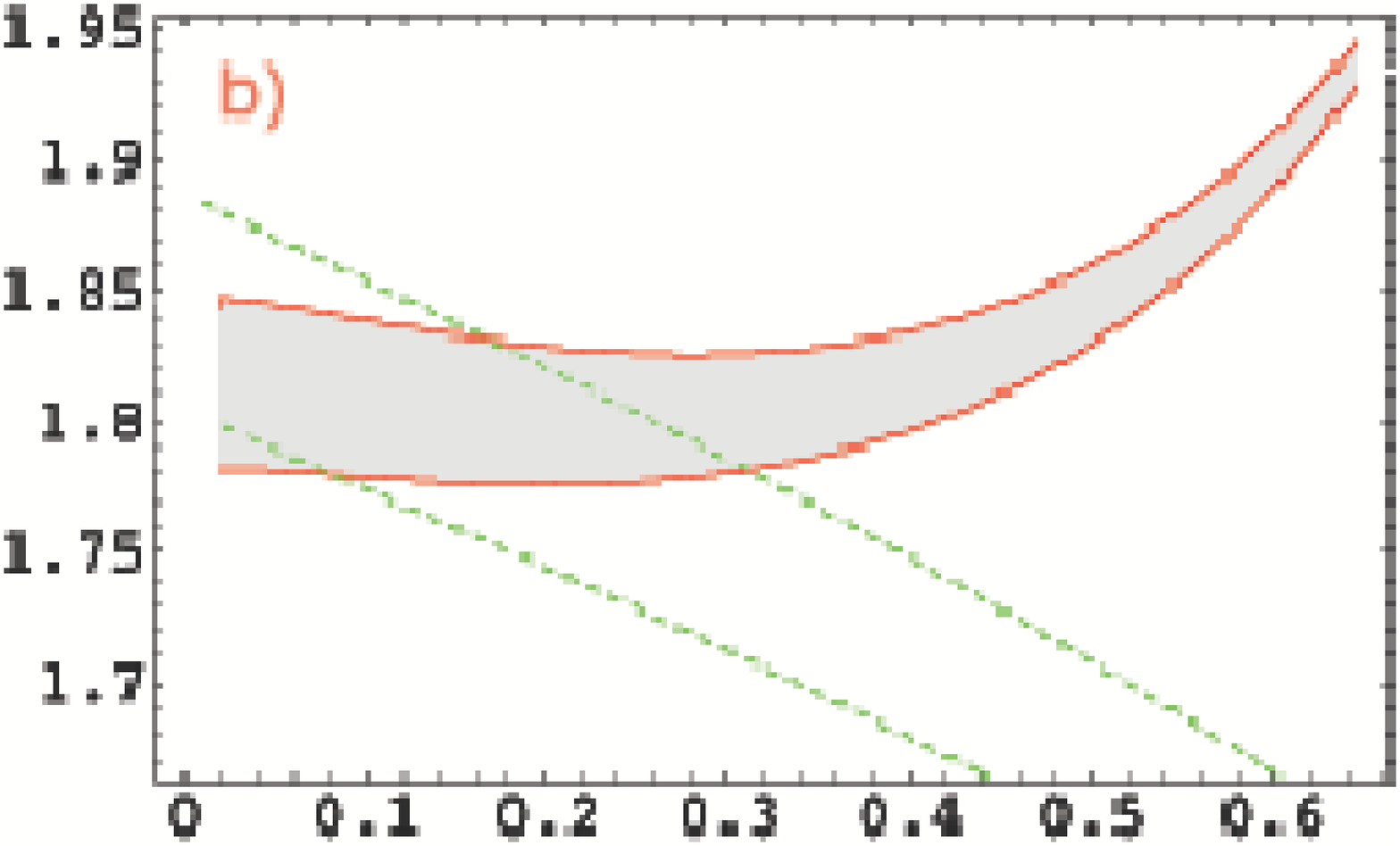}}
 \caption{\scriptsize 
Predictions of the LSR in the $1^{-+}$ hybrid channel using the central values of the QCD input parameters and two values of $t_c=4.6$ (beginning of the $\tau$-stability) and 5 GeV$^2$ fixed from the previous  FESR results: a) RG invariant coupling constant $\hat f_H$ from ${\cal L}_0$ (red: continuous), from ${\cal L}_1$ (black: dashed-dotted (on top of the red curve)), from ${\cal L}_1$ (green: dashed); b) $M_H$ from ${\cal R}_0(\tau)$ (red: continuous) and  from ${\cal R}_1(\tau)$ (green: dotted).
 }
\label{fig:lsr}
\end{figure}}
\end{center}
\vspace*{-0.5cm}
We show in Fig. \ref{fig:lsr} the predictions of the LSR in the $1^{-+}$ channel using the central values of the QCD input parameters and using two values of $t_c=4.6$ (beginning of the $\tau$-stability) and 5 GeV$^2$ fixed from the previous  FESR results.
The prediction for the decay constant is not conclusive as there is not a stability in the LSR variable $\tau$. However, on can see from Fig. \ref{fig:lsr} that the different predictions interact for :
\beq
\hat f_H= (10\sim 22)~{\rm MeV}~,
\eeq
which is consistent with the previous FESR result but less accurate. Therefore, we consider as a final prediction the one from FESR. 
For the mass prediction, the LSR gives:
\beq
M_H=1.81(3)(7)~{\rm GeV}:t_c=(4.6\sim 5.0)~{\rm GeV}^2,
\eeq
where the 1st and 2nd errors come respectively from $t_c$ and the QCD input parameters.  We take the (na\"\i ve arithmetic) average of the FESR and LSR results and we take the quadratic average of the errors. Then, we obtain:
\beq
\la M_H\ra =1.81(6)~{\rm GeV}~.
\label{eq:hmass}
\eeq
\section{Comparison with existing theoretical results}
\nin
The central value of the result in Eq. (\ref{eq:hmass}) has increased by $100-200$ MeV compared to the previous QSSR results quoted in Eq. (\ref{eq:qssr}). The difference with the one in Ref. \cite{LNP,CHET} is due to the inclusion of radiative corrections for both perturbative and power corrections here, which increase the mass prediction by about 200 MeV. The difference with Ref. \cite{KORNER} is the non-inclusion of the tachyonic gluon mass and the use of factorization for the four-quark condensate in \cite{KORNER}. In \cite{YANG}, a value of 1.2 GeV for the mass has been obtained \footnote{Also, in \cite{YANG} an operator having high number of $\gamma$ matrices has been considered in the light-cone gauge but its correspondence in a covariant gauge is not quite transparent. The current  looks to a be a tensor current while  from the definition of the matrix element,   the $1^{-+}$ hybrid meson contributes through its longitudinal (i.e. spin zero) part.}, which the author attributes to be due to the hybrid operator anomalous dimension. In this paper, we show that the effect of perturbative  radiative corrections including the one due to the anomalous dimension only decreases the mass predictions by 50 MeV.  This result is (a priori) expected where it is easy to show that perturbative radiative corrections tend to cancel in the ratio of moments used to extract the meson mass. A similar explicit example has been studied for the case of $2^{++} \bar qq$ current \cite{BAGAN}. The present result is slightly lower than  the lattice results \cite{LATT1,LATT2} quoted in Eq. (\ref{eq:latt}). As the $\alpha_s$ corrections are reasonnably small, we do not see in the QCD side any potential contributions which can restore the discrepancy between QSSR and lattice results. From the phenomenological side, a more involved parametrization of the spectral function might help, but in many known channels, the usual duality ansatz : one resonance + QCD continuum describes accurately the spectral function at the $t_c$ and $\tau$ stability points even for broad states like e.g. the $\sigma$ and $K^*_0$ mesons \cite{VENEZIA}.

\section{Masses of the $1^{-+}$ four-quark states}
\begin{center}
{\begin{figure}
{\includegraphics[width=6cm]{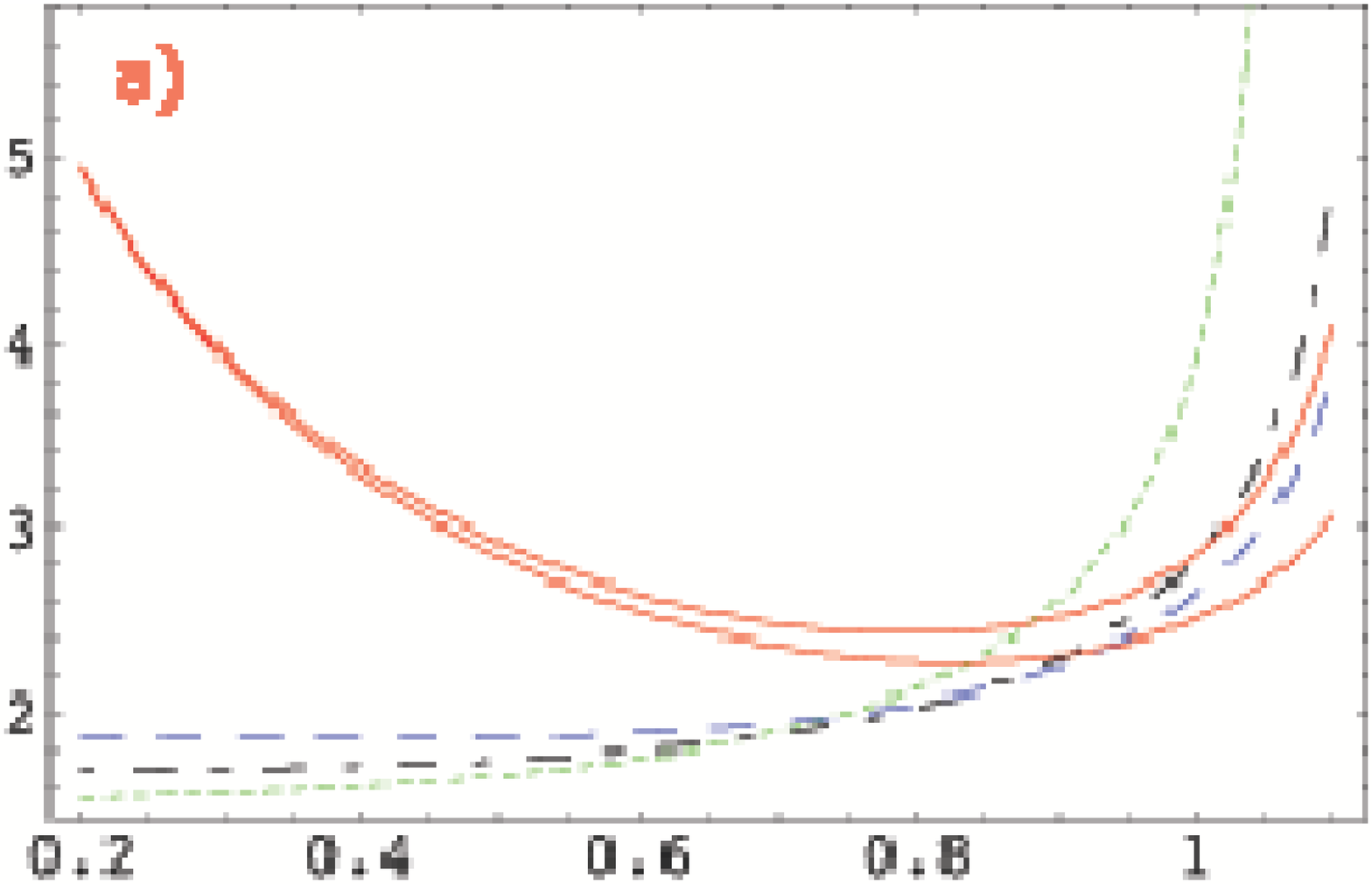}}
{\includegraphics[width=6cm]{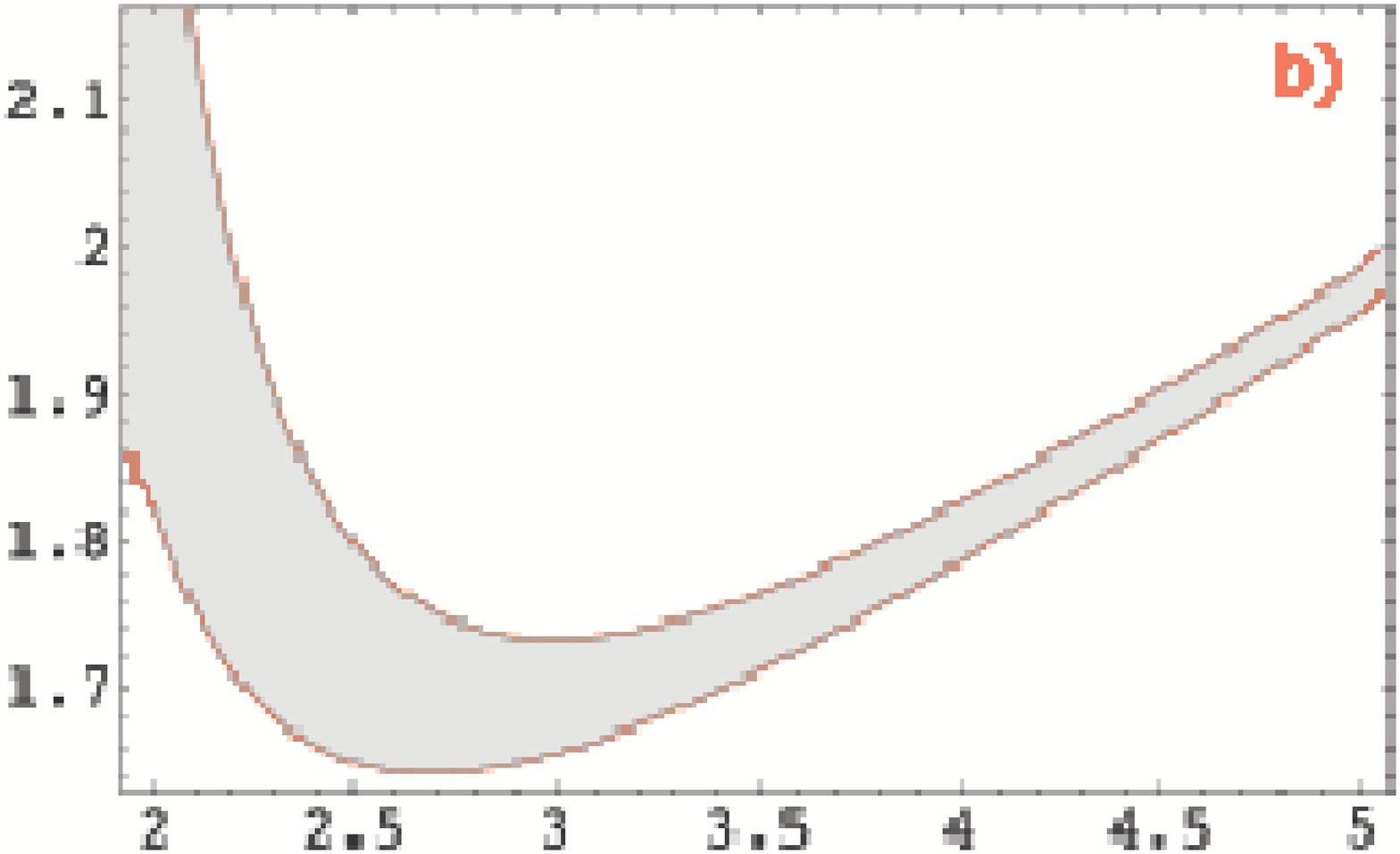}}
 \caption{\scriptsize 
Predictions of  the four-quark state mass in GeV using the current in Eq. (\ref{eq:4q1}): a) LSR: domain spanned by the upper bound versus $\tau$ using the largest range spanned by the QCD parameters given after Eq. (\ref{lamb}) (red: continuous curve); Mass using the central values of QCD parameters for $t_c=3 ~GeV^2$ (green: dashed), $t_c=4 ~GeV^2$ (black: dotted-dashed), $t_c=5 ~GeV^2$ (blue: dashed).  b) FESR versus $t_c$: region spanned by the largest range of QCD parameters.}
\label{fig:4q1}
\end{figure}}
\end{center}
\vspace*{-0.5cm}
\nin
We reconsider the QSSR analysis in \cite{ZHU} using the previous set of QCD parameters where the values of the gluon and four-quark condensates are about a factor 2 higher here. We shall introduce  the renormalization group invariant $\la \bar\psi\psi\ra$ condensate \cite{FNR}:
\beq
\hat\mu_i^3= {\la \bar\psi_i\psi_i\ra(\nu)\over \ga\log{\nu/\Lambda}\dr^{2/-\beta_1}}~,
\eeq
with \cite{SNmass}:
$
\hat\mu_u=-(263\pm 7)~{\rm MeV}~.
$
We neglect the small $Q^2$-dependence of $\alpha_s \la \bar\psi\psi\ra$ and $\la g\bar\psi G\psi\ra$ as 
well as the anomalous dimension of the four-quark current, which should have small effect in the mass determination. The four-quark meson  can be described by the diquark anti-diquark operators \footnote{In principle, these operators should mix under renormalizations \cite{TARRACH} and one should built their renormalization group invariant (RGI) combination for describing the physical state.}:
\bea
\eta_{1\mu}^M&=&u_a^TC\gamma_\mu d_b(\bar u_aC\bar d^T_b+\bar u_bC\bar d^T_a)+...\nnb\\
\eta_{3\mu}^M&=&u_a^TC d_b(\bar u_a\gamma_\mu C\bar d^T_b-\bar u_b\gamma_\mu C\bar d^T_a)+...
\label{eq:4q1}
\eea
where ... denotes an interchange between $\gamma_u$ and 1 or $\gamma_5$. The QCD expression of the LSR of the corresponding correlator can be written to LO in $\alpha_s$ \cite{ZHU}:
\bea
{\cal L}^{4q}(\tau)=&& \int_0^{t_c}dt \ga c_0t^4+c_4t^2+c_6t+c_8\dr e^{-t\tau}\nnb\\
&&+c_{10}+c_{12}\tau
\eea
where, for $\eta_{1\mu}^M$:
\bea
c_0&=&{1\over 18432\pi^6}~, ~~~~c_4=-{\la g^2 G^2\ra\over 18432\pi^6}~,\nnb\\
c_6&=&{\la\bar \psi\psi\ra^2\over 18\pi^2}~, ~~~~~~~c_8=-{\la\bar \psi\psi\ra\over 24\pi^2}\la g\bar \psi G\psi\ra,\nnb\\
c_{10}&=&{\la g\bar \psi G\psi\ra\over 192\pi^2}-{5\over 864\pi^2}\la g^2 G^2\ra\la\bar \psi\psi\ra^2,\nnb\\
c_{12}&=&-{32\over 81}g^2\la\bar \psi\psi\ra^4
+{\la g^2 G^2\ra\over 576\pi^2}\la\bar \psi\psi\ra\la g\bar \psi G\psi\ra~,
\eea
from which one can deduce the ratio ${\cal R}_n(\tau)$ and $R_n(t_c)$ defined in Eqs. (\ref{usr}) and (\ref{fesr}) used for extracting the meson mass. The result of the analysis is shown in Fig.~\ref{fig:4q1} from which we can deduce in units of GeV:
\beq
M_{4q1}\leq 2.4(1)~:~{\rm LSR}~~; ~~ M_{4q1}= 1.70(4)~:~{\rm FESR}~,
\label{eq:mass4q1}
\eeq
for $t_c\simeq (2.5\sim 3)$ GeV$^2. $
\begin{center}
{\begin{figure}
{\includegraphics[width=6cm]{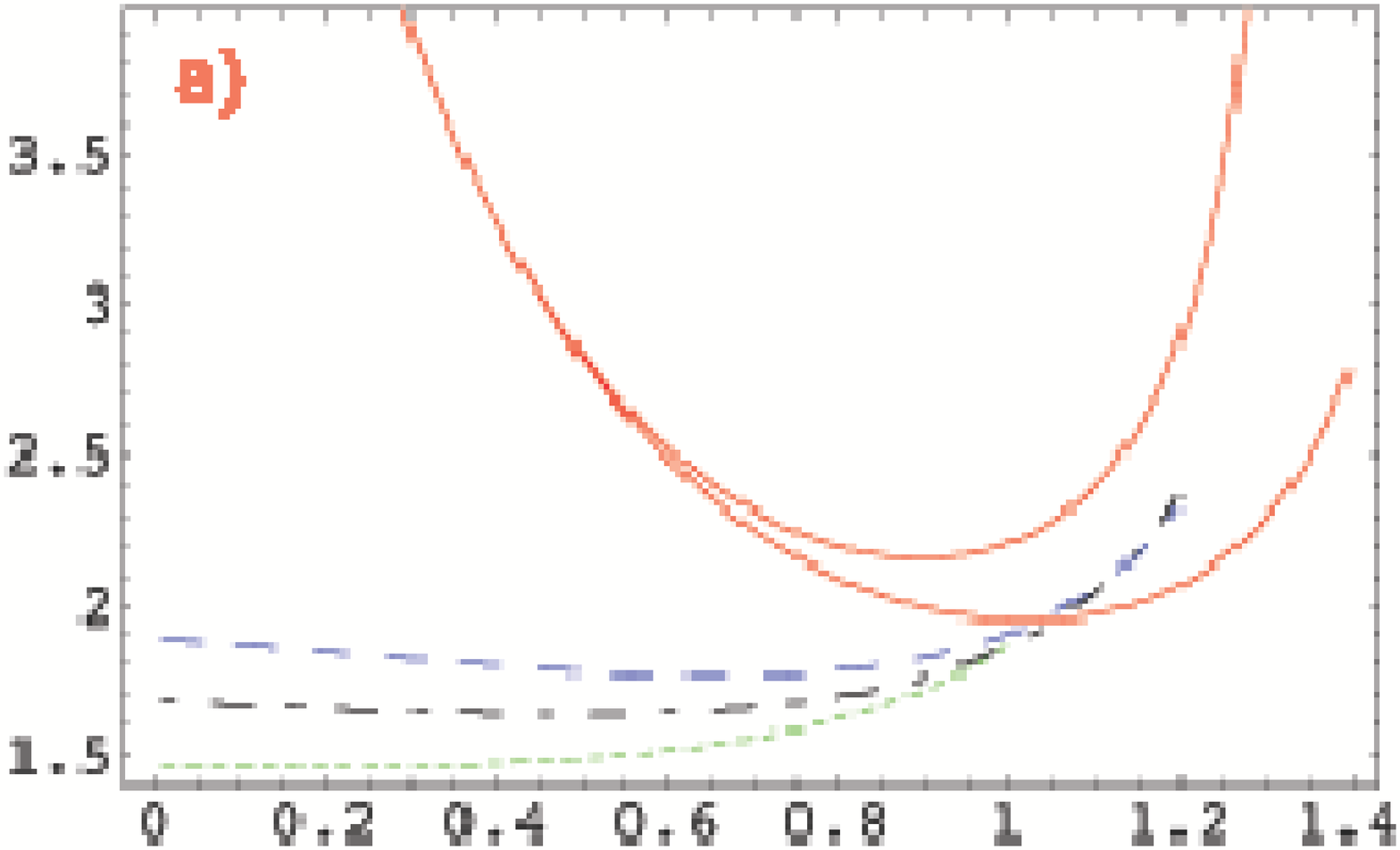}}
{\includegraphics[width=6cm]{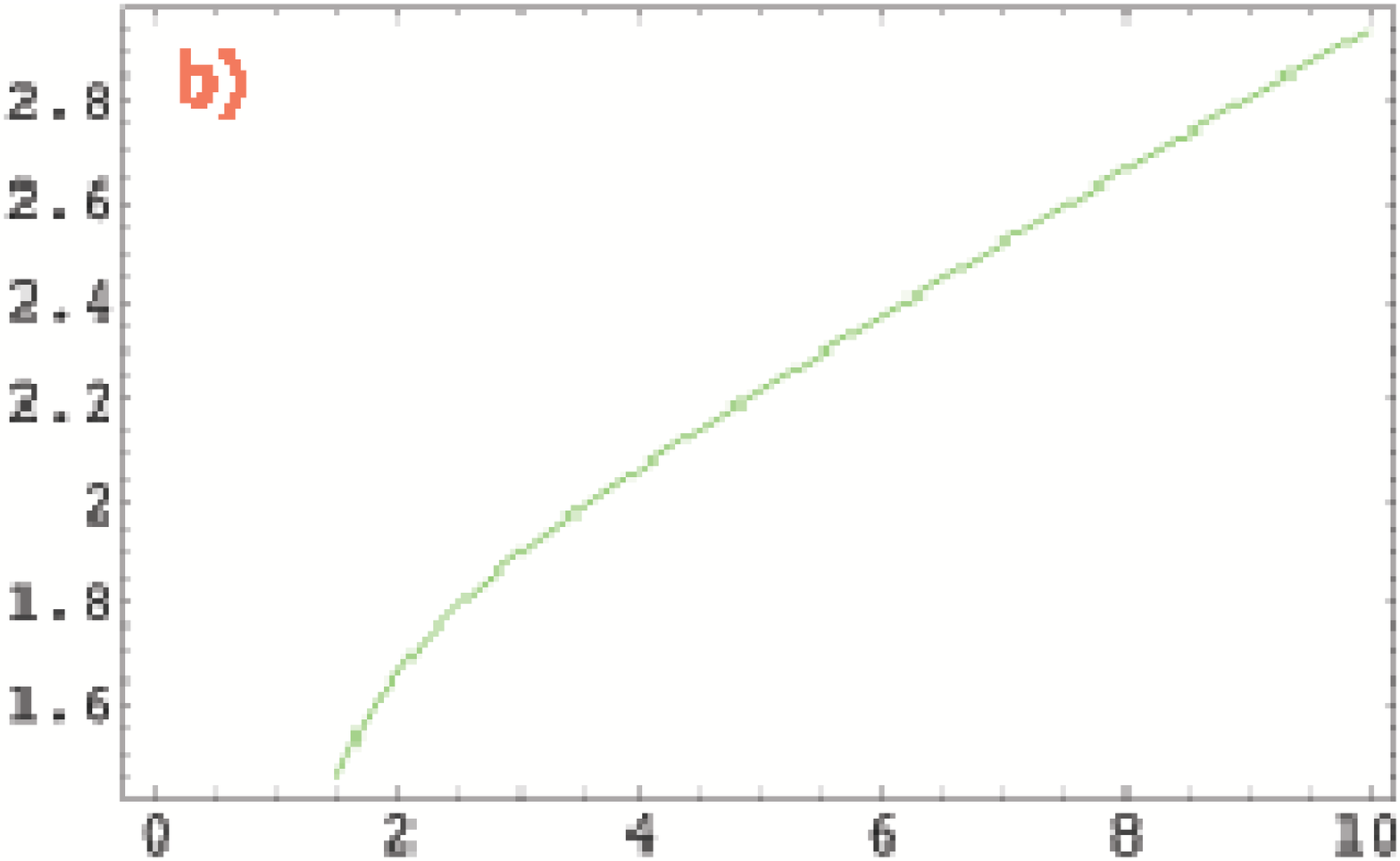}}
 \caption{\scriptsize 
Predictions of  the four-quark state mass in GeV using the current in Eq. (\ref{eq:4q3}): a) LSR: domain spanned by the upper bound versus $\tau$ using the largest range spanned by the QCD parameters given after Eq. (\ref{lamb}) (red: continuous curve); Mass using the central values of QCD parameters for $t_c=3 ~GeV^2$ (green: dashed), $t_c=4 ~GeV^2$ (black: dotted-dashed), $t_c=5 ~GeV^2$ (blue: dashed).  b) FESR versus $t_c$ using the central values of QCD parameters.}
\label{fig:4q3}
\end{figure}}
\end{center}
\vspace*{-0.5cm}
\nin
where one can notice that the LSR result does not have $\tau$- and $t_c$-stabilities such that only an upper bound can be extracted from the positivity of the spectral function. This result is slightly higher than the one in \cite{ZHU} due to the consideration of the violation of factorization in the estimate of the high-dimension condensates here. 
We have repeated the analysis for the $\eta_{3\mu}^M$.
We show in Fig \ref{fig:4q3}a) the result of the analysis from the LSR moment ${\cal R}_0(\tau)$ which presents $\tau$-stability for $t_c\geq 4$ GeV$^2$ but no $t_c$-stability. In the expanded form, FESR $R_0(t_c)$ presents a zero for $2.5\leq t_c\leq 7$ GeV$^2$ due to ${\cal M}_0(t_c )$ signaling large non-perturbative effects for this quantity. This zero shows up like a minimum in $t_c$ in the non-expanded form of  $R_0(t_c)$ given a mass $M\approx 1.2$ GeV, but at too low value of $t_c\approx 2. $ GeV$^2$, where the OPE does not converge \footnote{In \cite{ZHU}, this minimum corresponds to a slightly higher value of the mass due
to the different choice of the QCD set of parameters.}. At this $t_c$-values, the LSR does not also show a $\tau$-stability. Therefore, we do not retain this result from our analysis. $R_1(t_c)$ moment  has a much better behaviour but (unfortunately) does not show a $t_c$-stability [see Fig.  \ref{fig:4q3}b)]. The most conservative result from both LSR and FESR is:
\beq
1.64\leq M_{4q3}\leq 2.1(1)~:~~~t_c\geq 4~{\rm GeV}^2~.
\label{eq:mass4q3}
\eeq
where the upper bound comes from the positivity of the LSR moment.   Currents with higher number of $\gamma$ matrices have been also considered in \cite{ZHU}:
\beq
\eta_{2\mu,4\mu}^M\equiv u_a^TC\gamma^\nu\gamma_5 d_b(\bar u_aC\sigma_{\mu\nu}\gamma_5\bar d^T_b\pm \bar u_bC\sigma_{\mu\nu}\gamma_5 \bar d^T_a+...)~,
\label{eq:4q3}
\eeq
where ... is an interchange between $\gamma^\nu$ and $\sigma_{\mu\nu}$. The analysis of the corresponding correlators lead to similar results (within the errors) than the ones in Eq. (\ref{eq:4q1}):  
\beq
M_{\eta_{4\mu}}\approx M_{\eta_{1\mu}}~~~{\rm  and}~~~ M_{\eta_{3\mu}}\approx M_{\eta_{2\mu}}~.
\eeq
For definiteness, we consider that we have only one 4-quark state coupled to a RGI operator which results from a combination of these different operators expected to mix under renormalizations \cite{TARRACH}. We fix the four-quark state mass to the value from Eq. (\ref{eq:mass4q1}):
\beq
M_{4q}= 1.70(4)~{\rm GeV}~.
\label{eq:mass4q}
\eeq 
\section{Masses of the $1^{-+}$ molecules}
\begin{center}
{\begin{figure}
{\includegraphics[width=7cm]{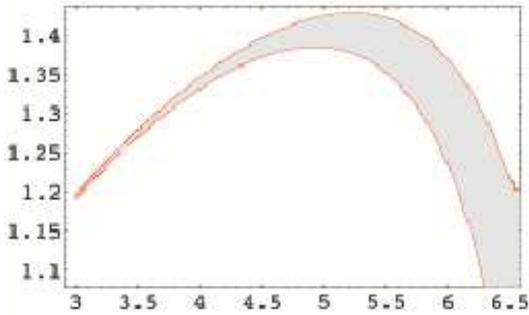}}
 \caption{\scriptsize 
Mass of the molecule state associated to the current $J_1^\mu$ in Eq. (\ref{eq:mol}) versus $t_c$ using the largest range spanned by the QCD parameters given after Eq. (\ref{lamb}) (red: continuous curve).}
\label{fig:mol1}
\end{figure}}
\end{center}
\vspace*{-0.5cm}
\nin
The possibility to form  two $1^{-+}$ molecules with the two operators:
\bea
J_1^\mu&\sim&(\bar\psi\gamma_5\psi)(\bar\psi\gamma_5\gamma^\mu\psi)~,\nnb\\
J_2^{\mu\nu}&\sim&\epsilon^{\nu\nu\rho\sigma}\Big{[}(\bar u\gamma_5\gamma_\rho d)(\bar d \gamma^\sigma u)
-(\bar d\gamma_5\gamma_\rho u)(\bar u \gamma^\sigma d)\Big{]},
\label{eq:mol}
\eea
has been discussed in \cite{ZHANG}, where the meson associated to the 1st (resp 2nd) operator is expected to decay into $\eta\pi,~\eta'\pi$ (resp. $\rho\pi,~b_1\pi$). Using the QCD expression given in \cite{ZHANG} without the one-direct instanton contribution which is largely affected by the uncertainty
of the instanton density $\rho$ appearing as $\rho^6$, we use the lowest FESR moment $R_0(t_c)$ for the analysis of the meson mass $M_{mol1}$ associated to the current $J_1^\mu$ where a $t_c$-stability is obtained at $t_c\simeq 5$ GeV$^2$ (Fig. \ref{fig:mol1}), while the LSR moments ${\cal R}_{0,1,2}$ do not give conclusive results. The mass $M_{mol2}$ associated to the current $J_2^\mu$ increases with $t_c$ for the FESR, while the LSR moments decrease slowly with the sum rule variable $\tau$ but present a zero around $\tau = 0.7-0.8$ GeV$^{-2}$ (not shown in Fig. \ref{fig:mol2}), while it 
presents $t_c$-stability around 3 GeV$^2$. The optimal results are:
\beq
M_{mol1}=1.41(3)~{\rm GeV}~,~~~~~~M_{mol2}\simeq (1.2\sim 1.4){\rm GeV}~,
\label{eq:massmol1}
\eeq
which agree within the errors with the ones in \cite{ZHANG}. However, we can consider that
the two operators in Eq. (\ref{eq:mol}) mix under renormalization such that only one physical state couples to the corresponding RGI operator, which we fix to be:
\beq
M_{mol}=1.3(1)~{\rm GeV}~.
\label{eq:massmol}
\eeq
\begin{center}
{\begin{figure}
{\includegraphics[width=7cm]{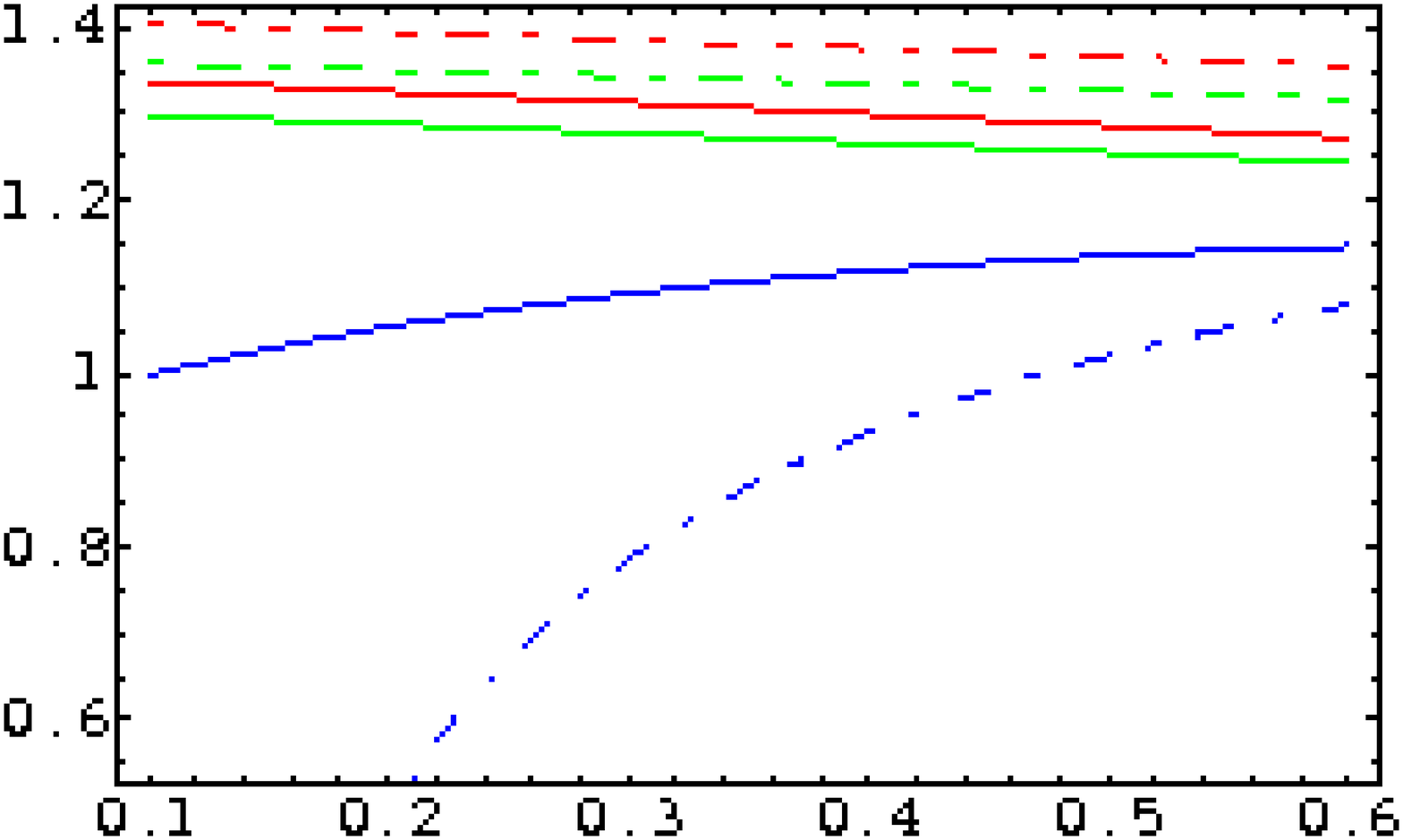}}
{\includegraphics[width=7cm]{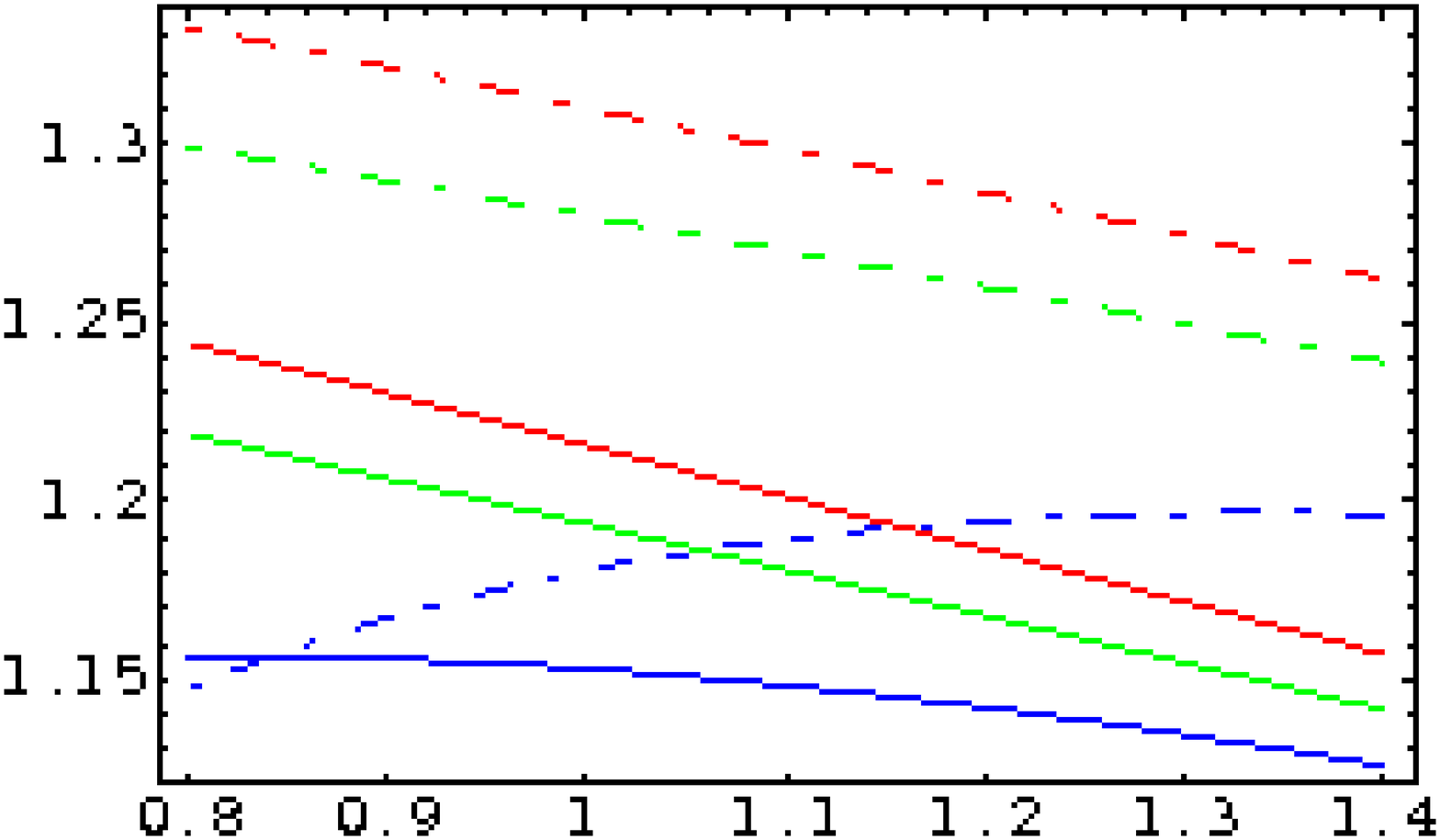}}
 \caption{\scriptsize 
Mass of the molecule state associated to the current $J_2^\mu$ in Eq. (\ref{eq:mol}) from LSR versus the sum rule variable $\tau$= 0.1 to 0.6 GeV$^{-2}$ and 0.8 to 1.4 GeV$^{-2}$ and for different values of  $t_c$ using the central values of the QCD parameters given after Eq. (\ref{lamb}) 
and for two LSR moments ${\cal R}_0$ (continuous curve) and ${\cal R}_1$ (dotted-dashed curve):
$t_c=2.5$ GeV$^2$ (green), $t_c=3$ GeV$^2$ ( red) and $t_c=3.5$ GeV$^2$ (blue).}
\label{fig:mol2}
\end{figure}}
\end{center}
\vspace*{-0.5cm}
\nin
\section{Nature of the $\pi_1(1400)$, $\pi_1(1600)$ and $\pi_1(2015)$}
\nin
From your previous analysis and for a further phenomelogical use, we shall consider one hybrid, one four-quark and one molecule states below 2 GeV, with the masses from Eqs. (\ref{eq:hmass}), (\ref{eq:mass4q}) and (\ref{eq:massmol}) in GeV:
\beq
M_H=1.81(6)~,~M_{4q}=1.70(4)~, ~M_{mol}=1.3(1)~.
\label{eq:mass}
\eeq
The $\pi_1(1400)$ is intriguing as it is seen to decay into $\eta\pi$ and $\eta'\pi$ but not into $\rho\pi,~b_1\pi$ \cite{PDG} \footnote{For a review on different experimental results and related problems, see e.g. \cite{KLEMPT}.} . The later decays being also expected for an hybrid state \cite{LNP}. Looking at our {\it bare (unmixed)} mass predictions in Eq (\ref{eq:mass}), one may expect that the molecule state is the most probable candidate. We consider that the suppression of the $\rho\pi$ decay can be due to a mixing of this molecular state with the four-quark or/and hybrid states. We use a minimal  two-component mixing:
\bea
\pi_1(1400)&=& \cos\theta_{mol} |mol\ra +\sin\theta_{mol} |X\ra\nnb\\
\pi_1(1600)&=& -\sin\theta_{mol} |mol\ra +\cos\theta_{mol} |X\ra~,
\eea
where $X$ is a four-quark or hybrid state. The ``best fit" is obtained for the sets : 
\bea
(M_{mol},M_X)&=&(1.2\sim 1.3~;~1.70,1.74)~{\rm GeV} \lrar\nnb\\
 \theta_{mol}&\simeq& -(11.7\pm 2.2)^0~,
\eea
which is slightly favours a molecule/four-quark mixing. This
result may suggest that the  $\pi_1(2015)$, quoted in the extended version of PDG \cite{PDG}, can have more hybrid state in its wave function. 
One also expects that the non-exotic meson $1^{--}$ as well as the isospin partners  of these $1^{-+}$ exotics should be almost degenerate in masses with these exotic states, while some of their radial excitations have  masses of the order of optimal continuum threshold: 
\beq
M'\approx \sqrt{t_c}\simeq (1.7\sim 2.2)~{\rm GeV}~. 
\eeq 
Further experimental ad theoretical tests of this mixing scheme are required.
\vfill\eject
\section*{Acknowledgement}
\nin
This work has been partially  supported by the CNRS-IN2P3 within the French-China Particle Physics Laboratory (FCPPL) for  the visit at IHEP (Beijing) and at the University of Nankai (Tianjin)  and by Peiking University  (Beijing). 


\end{document}